\documentclass[sigconf]{acmart}
\AtBeginDocument{%
  \providecommand\BibTeX{{%
    \normalfont B\kern-0.5em{\scshape i\kern-0.25em b}\kern-0.8em\TeX}}}

\setcopyright{acmcopyright}
\copyrightyear{2018}
\acmYear{2018}
\acmDOI{XXXXXXX.XXXXXXX}

\acmConference[Conference acronym 'XX]{Make sure to enter the correct
  conference title from your rights confirmation emai}{June 03--05,
  2018}{Woodstock, NY}
\acmPrice{15.00}
\acmISBN{978-1-4503-XXXX-X/18/06}
\usepackage{caption}
\usepackage{subcaption}
\begin{document}

\title{Model-agnostic Counterfactual Synthesis Policy for Interactive Recommendation}

\author{Siyu Wang}
\affiliation{%
  \institution{University of New South Wales}
  \city{Sydney}
  \state{NSW}
  \country{Australia}
}
\email{siyu.wang5@student.unsw.edu.au}

\author{Xiaocong Chen}
\affiliation{%
  \institution{University of New South Wales}
  \city{Sydney}
  \state{NSW}
  \country{Australia}
  }
\email{xiaocong.chen@unsw.edu.au}

\author{Lina Yao}
\affiliation{%
  \institution{University of New South Wales}
  \city{Sydney}
  \state{NSW}
  \country{Australia}
}
\email{lina.yao@unsw.edu.au}

\renewcommand{\shortauthors}{Trovato and Tobin, et al.}

\begin{abstract}
Interactive recommendation is able to learn from the interactive processes between users and systems to confront the dynamic interests of users. Recent advances have convinced that the ability of reinforcement learning to handle the dynamic process can be effectively applied in the interactive recommendation. However, the sparsity of interactive data may hamper the performance of the system. We propose to train a Model-agnostic Counterfactual Synthesis Policy to generate counterfactual data and address the data sparsity problem by modelling from observation and counterfactual distribution. The proposed policy can identify and replace the trivial components for any state in the training process with other agents, which can be deployed in any RL-based algorithm. The experimental results demonstrate the effectiveness and generality of our proposed policy.

\end{abstract}





\maketitle

\section{Introduction}
Recommendation systems have been increasingly used by the industry, such as social media, e-commerce and digital streaming, to provide users with personalised content.
Traditional recommendations are typically developed using content-based filtering or collaborative filtering approaches~\cite{ramlatchan2018survey}, which predict the user's future interest based on past preference. However, since the preference of the user can change over time, modelling from past interests may not give an accurate prediction.
In order to understand the changing interest of users, the interactive recommendation was introduced as a practical way to improve recommendation systems due to supporting an interactive process~\cite{mahmood2009improving}. The interactive recommendation is a process in which the system takes an optimal action in each step to maximise the user's feedback reward. Since Reinforcement Learning (RL) can learn from users' instant feedback, it has been regarded as having the ability to confront the interactive process. Therefore, reinforcement learning has been considered an effective method to model interactive recommendation system~\cite{dulac2015deep, zheng2018drn,chen2021survey}.

Although significant efforts have been made in this field, there are still some challenges remaining to be further studied. 
In general, users cannot interact with all items in the recommendation system, resulting in missing reviewing data for a portion of items that have not been interacted with.
For this reason, only a fraction of the interaction data between users and items can be gathered by the recommendation system. 
However, it is difficult for the system to understand users' preferences and make satisfactory recommendations without having a substantial amount of interaction data.
When using the RL to achieve a recommendation system, the agent will take action depending on the current state of the environment, interact with the environment and receive a reward.
However, if there is no record of this situation, the system will mistakenly believe that the user is not interested in the item and return a zero reward.
In this case, the recommendation system may fail to capture the actual preferences of users.
The sparsity interaction between users and items might restrict the ability of the system and destroy the satisfaction of users.
Therefore, data sparsity is still a challenge confronted by the recommendation system.

% we can only get limited interactions.
% items can not be interacted 
% data sparsity -> challenge 
% To address this challenge, Recent work has made several initial attempts to relief sparsity by using data augmentation. Specifically, causality is used, xxx ...., xxxx ....... However, t

% Compared with the number of items in a system, which can easily reach millions, a user can only view a fraction of the items where a more mere portion can get the feedback of the user. The sparsity interaction between users and items might restrict the ability to learning a accurate and efficient system.

To address this challenge, recent works have made several initial attempts to relieve sparsity by using data augmentation.
Specifically, causality, as a critical research topic, has been used in recommendation systems to address this problem.
% Since the causality can analyze the response to the occurrence of an effect based on a causal connection, the counterfactual data can be regarded as an answer to a counterfactual question, "what the interaction process would be if we intervene on the observed data?"
~\citet{zhang2021causerec} propose to identify dispensable and indispensable items in behaviour sequence by measuring the similarity between the representation of each item and target item and leveraging the attention mechanism as well. By replacing the top half items with the lowest similarity scores, they can obtain the positive counterfactual user sequence.
~\citet{wang2021counterfactual} design a sampler model that makes small changes to the user's historical items on the embedding space to implement a counterfactual data augment framework. However, these approaches are all based on the embedding space, which is not available in the reinforcement learning environment. Although the state representation has similarities to the embedding, they are conceptually different. State representation in RL-based recommendation system consists of users' recent interest and demographic information, while traditional embedding method only contains users' recent actions. In addition, the state representation is dynamic, which is affected by the action while the embedding is fixed. Therefore, existing frameworks are not suitable for RL-based interactive recommendations in response to the changing interest of users.

On the other hand, the causality can analyze the response to the occurrence of an effect based on a causal connection, the counterfactual data can be regarded as an answer to a counterfactual question, "what the interaction process would be if we intervene on the observed data?". Depending on the observed data, the generated counterfactual data can be considered as complementing the situation not covered by the observed data and thus is a powerful data augmentation tool to help the recommendation system understand users' real interests. Motivated by the above, we aim to leverage counterfactual in the RL-based interactive recommendation to address the data sparsity problem. To enhance the generalization ability, we propose an approach to learn a model-agnostic counterfactual synthesis policy to find a counterfactual for a state where the trivial components have been replaced.

% Based on the above, we aim to make it possible for a policy to generate counterfactual interaction data that can take into account the dynamic preference of the user. 
% To this end, we propose an approach to learn a counterfactual synthesis policy to find a counterfactual for a state where the trivial components have been replaced. 
Specifically, we identify the trivial components and infer the causal mechanism by minimising the difference between the rewards that pre-trained policy received from the environment under original and counterfactual state. We treat the state with minimal reward difference as the counterfactual for the original one where the trivial components have been replaced. The counterfactual synthesis policy can be deployed into any RL frameworks and generate counterfactual interactive data along with other agent during interaction. Through modelling from both counterfactual and observation interaction data, the system can confront the data sparsity problem effectively.
Our main contributions are summarized as:
\begin{itemize}
    \item We propose to address the data scarcity problem in the interactive recommendation by modelling both observation and counterfactual distribution.
    % \item We design a novel counterfactual synthesis policy, which can be deployed in different RL-based frameworks to generate counterfactual interaction data automatically.
    \item We design a novel model-agnostic counterfactual synthesis policy for the interactive recommendation, which can capture the changing interest of users and be easily deployed in various RL-based frameworks to generate counterfactual interaction data along with the training agent.
    \item We conduct extensive experiments and show that applying counterfactual synthesis policy on RL-based framework can significantly improve the performance of interactive recommendation.
\end{itemize}

\section{Model-agnostic Counterfactual Synthesis Policy}

\begin{figure}[h]
  \centering
  \includegraphics[width=\linewidth]{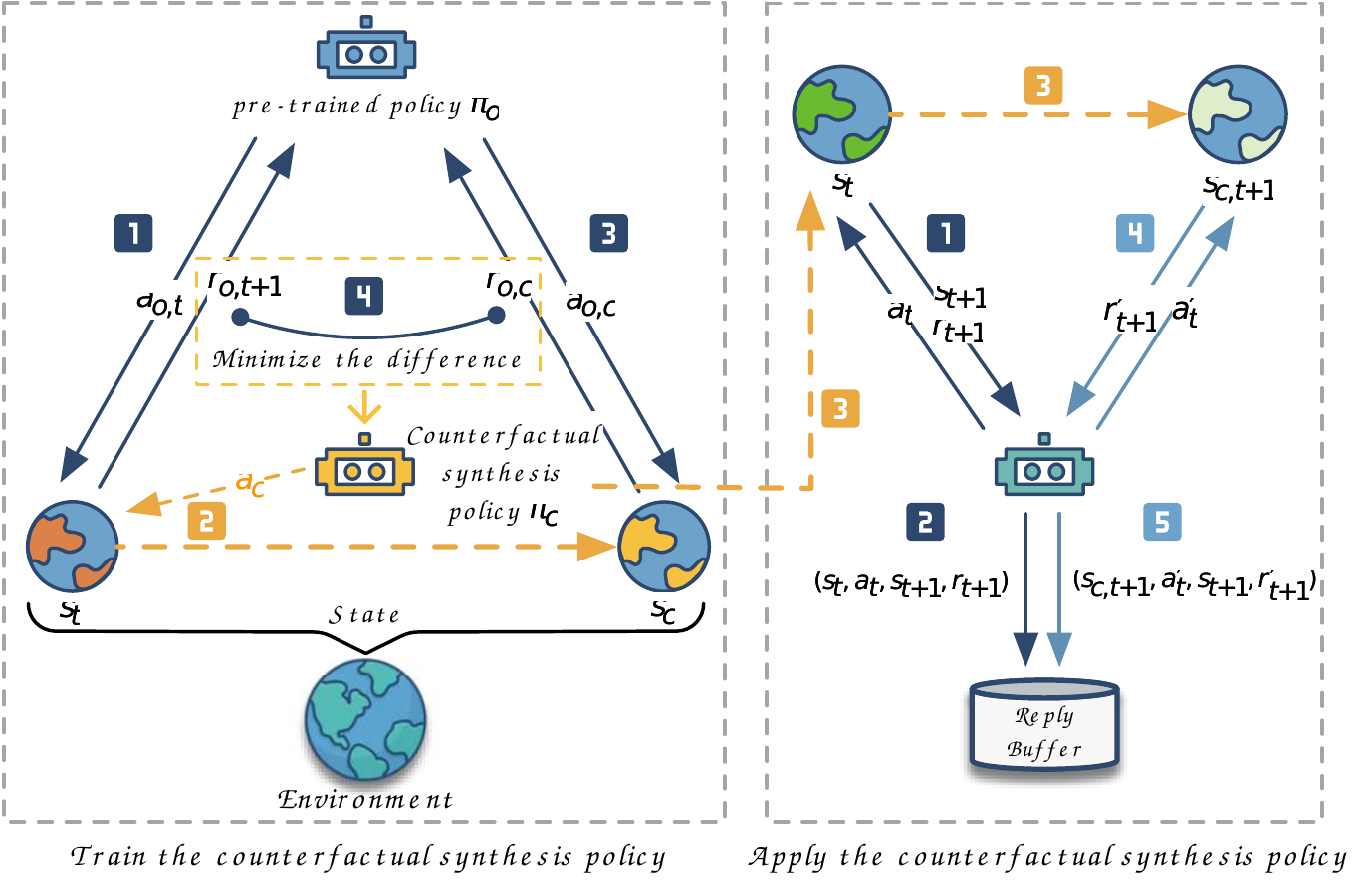}
  \caption{An illustration of proposed counterfactual synthesis policy. The left part is the schematic of how to train the counterfactual synthesis policy. The right part is how to apply the counterfactual synthesis policy to other RL algorithms for data augmentation.}
  \Description{.}
\end{figure}

From the perspective of causal reasoning, the data generation process of observed data may be influenced by multiple causes. The essence of our counterfactual synthesis policy is to find a subset of causes and perform interventions on it to obtain counterfactual. We introduce two concepts of the causes in the data generation process, essential components and trivial components, to identify which subset we should make the interventions.

Among the set of causes that can be manipulated, the degrees of causal effects are different among different causes. Some causes have considerable influence on the interactive process of generating data, which is related to a meaningful aspect of the user's interest. We refer to these parts of causes as essential components. In contrast, the trivial components represent a subset of causes having less influence on the interactive process, which may contain more noisy terms

% The policy is designed to generate the counterfactual by performing interventions on the trivial components. In this way, the counterfactual data would have a similar semantic with the observed one, resulting minimal causal effects on the user’s interest. Therefore, we construct reasonable counterfactual data distributions that differ from the observational distribution to handle the issue of sparse interaction in the RL-based interactive recommendation. Modelling both the observation and counterfactual distribution will also make the system less sensitive to the trivial components.

Given the identified two concepts, the policy is designed to generate the counterfactual by performing interventions on the trivial components. In this way, the counterfactual data would have a similar semantic with the observed one, resulting in minimal causal effects on the user’s interest. The main tasks of learning the counterfactual synthesis policy involve estimating the causal effects, finding the trivial components and performing interventions. 
The achievement of these goals relies on the investigation of the causal relationships during the data generation process. To this end, we introduce the Structural Causal Model (SCM), which is a critical part of formalizing causal reasoning and causal learning.
With a directed acyclic graph (DAG) \( \mathcal{G} \), 
SCM $\mathcal{M}$ consists of a set of structural equations ~\cite{peters2017elements}:
\begin{equation}
\label{eqt-l-inf}
\begin{aligned}
Y_i := f_i(V_i, U_i), \:\: i = 1,...,n, 
\end{aligned}
\end{equation}
where:
\begin{itemize}
    \item $Y = \{Y_1,...,Y_n\}$ is a set of observed variables, called effect. And $V = \{V_1,...,V_n\}$ is a set of parents of $Y$, representing a direct cause of $Y$.
    \item $U = \{U_1,...,U_n\}$ represents independent noise determined by omitted and unobserved factors. We assume that $U_i$ is independent of all other noise variables.
    \item $F = \{f_1,...,f_n\}$  is a set of structural equations. Each $f_i$ is a causal mechanism that determines the value of $Y_i$ based on the values of $V_i$ and the noise term $U_i$.
\end{itemize}

% The main tasks of learning a counterfactual synthesis policy involve estimating the causal effects, finding the trivial components and performing interventions. To achieve these goals efficiently, we propose to leverage the SCM in Reinforcement Learning to learn a counterfactual synthesis policy. We assume that the state $S_{t+1}$ satisfies the following SCM~\cite{lu2020sample}:

We can obtain intervention distributions and counterfactuals with SCM by estimating the system behaviour under intervention. The common ways to do the intervention contain making interventions on one (or several) variable(s), replacing one (or several) structural equation (s) or the distribution of noise variables~\cite{peters2017elements}. In general, the intervention distributions we obtained by performing interventions differ from the observational distribution.

Specifically, we leverage SCM in Reinforcement Learning to infer the causality and identify the trival components. Assume that the state $S_{t+1}$ satisfies the following SCM~\cite{lu2020sample}:
\begin{equation}
\label{eqt-l-inf}
\begin{aligned}
S_{t+1} = f_t(S_t, A_t, U_{t+1}),
\end{aligned}
\end{equation}

where $A_t$ is the action taken by the RL agent at time $t$. Depending on the current state $S_t$ of the environment and the action $A_t$, the environment goes into the next state $S_{t+1}$ according to the Markov Decision Process.

Since the action variables $A$ influence the next state $S_{t+1}$, we can estimate the causal effects under intervention on the action. 
A basic method is using mathematical do-calculus to simulate interventions by replacing the value of a variable $A$ with $a_c$, 
% \textcolor{red}{
% We use the basic method, mathematical do-calculus, to simulate interventions by replacing the value of a variable $A$ with $a_c$,
% }
donated as $do(A := a_c)$.
Then the intervention distribution $p^{~c}_{do(A :=a_c)}(s_{t+1})$ is identifiable~\cite{peters2017elements}:
\begin{equation}
\label{eqt-l-inf}
\begin{aligned}
p^{~c}_{do(A :=a_c)}(s_{t+1}) = p^{c}(s_t, \widetilde a, u_{t+1})p^{c}(s_{t+1}|s_t, \widetilde a, u_{t+1})
\end{aligned}
\end{equation}

% The main tasks of learning a counterfactual synthesis policy involve estimating the causal effects of causes and finding the subset that have minimal causal effects on the user’s interest. To achieve the both goals efficiently, we propose to use Reinforcement Learning to learn a policy, which can automatically find the subset of causes that have minimal causal effects and generate counterfactual interaction data.
Hence, we can construct reasonable counterfactual data distributions that differ from the observational distribution to handle the issue of sparse interaction in the RL-based interactive recommendation. Modelling both the observation and counterfactual distribution will also make the system less sensitive to the trivial components. Different from existing work that focuses on the estimation of causal mechanism $f_i$~\cite{lu2020sample}, we infer the causal mechanism during the learning of policy and estimate the causal effects based on the rewards of the environment. With a pre-trained policy $\pi_o$ from the observation distribution, our policy $\pi_c$ is trained to replace trivial components for any state and generate counterfactuals. 

% \begin{figure}[h]
%   \centering
%   \includegraphics[width=\linewidth]{fig1.eps}
%   \caption{An illustration of proposed counterfactual synthesis policy. The left part is the schematic of how to train the counterfactual synthesis policy. The right part is how to apply the counterfactual synthesis policy to other RL algorithms for data augmentation.}
%   \Description{.}
% \end{figure}

The training process of our counterfactual synthesis policy is illustrated in the left part of Figure 1.
Depending on the current state $s_t$ at time $t$, the agent following the pre-trained policy $\pi_o$ takes the action $a_{o, t}$ and interacts with the environment to obtain the reward $r_{o, t+1}$. 
To simulate intervention, we apply do-calculus under the same environment whose current state is $s_t$. By replacing the action $a_{o, t}$ with the action $a_c$ taken by our training agent, the environment would enter into the counterfactual state $s_c$.
% When facing the same environment with the current state $s_t$, the environment would enter into the counterfactual next state $s_{c, t+1}$ based on the action $a_c$ chosen by policy $\pi_c$.
Then we estimate the causal effect by putting the agent with pre-trained policy $\pi_o$ in the environment, of which the current state is counterfactual $s_c$, to receive the reward $r_{o,c}$.

Since the agent is required to be trained to identify and replace the trivial components, we measure the difference between the reward $r_{o, c}$ and $r_{o, t+1}$ and expect to minimize the difference. The difference indicates whether the state $s_c$ is a counterfactual for state $s_t$ where the trivial components have been replaced. If the agent with pre-trained policy $\pi_o$ takes similar actions in two environments (current state are $s_t$ and $s_c$, respectively) and receives similar rewards, the essential components of states $s_t$ and $s_c$ remain unchanged. Only the trivial components that contain more noise have been replaced. In contrast, the larger difference means a deviation from the counterfactual state to the original state, which may change the key aspect of user interest.

% It can be used for any DRL algorithms, xxxxxxxx.

Deep Deterministic Policy Gradient(DDPG) is a model-free and off-policy algorithm based on actor-critic architecture,  which uses deep function approximators that can learn policies in an environment with high-dimensional and continuous action spaces~\cite{lillicrap2015continuous}. 
Along with the above features and its simplicity to implement and scale to difficult problems, we apply the DDPG to train our policy~\cite{chen2020knowledge} and the training objective can be indicated as minimising the loss function:
\begin{align}
L(\theta_\mu, \mathcal{D}) = \mathop{E}_{(s, a, s', r) \sim \mathcal{D}} \Big[\Big(\big(r+\gamma(\mu_{\theta^{targ}_\mu}(s', \phi_{\theta^{targ}_\phi}(s'))\big) - \mu_{\theta_\mu}\big(s, a\big)  \Big)^2 \Big],
\end{align}
where $\mathcal{D}$ is a set of mini-batch of transitions $(s, a, s', r)$ for $s \in S$, $a\in A(s)$, $r \in R$, and $s' \in S^+$ ($S^+$ is $S$ plus a terminal state). $\theta_\mu$ and $\theta_\phi$ are parameters for critic and actor network, respectively. And $\mu_{\theta^{targ}_\mu}$ represent the target critic network.

% The pre-trained counterfactual synthesis policy can be used for any RL algorithm, and the brief illustration is depicted in the right part of Figure 2. 
% The pre-trained counterfactual synthesis policy can generate a counterfactual state during the interaction. 
% Put the training agent in the environment with the counterfactual state and record the action taken by the training agent. Then we can have a counterfactual transition, including counterfactual state, counterfactual action, real next state and real reward. By storing the counterfactual transition into the replay memory buffer, we can achieve the counterfactual generation through the training process.
% Original trained policy π

% s_t→π→a_t:
% Interact with the env to obtain r_{t+1} 

% OUR POLICY π_c
% Choose action a_c
% Interact with the env to obtain s_{t+1} 

% s_{t+1} → π→a_{t+1} 
% Interact with the original env(s=s_t) to obtain  r_{t+2} 
\section{Experiments}
\subsection{Experiments Setup}
\vspace{1mm}\noindent\textbf{Dataset.}
We conduct experiments and evaluate our model on VirtualTaobao~\cite{shi2019virtual}, a real-world online retail environment. 
VirtualTaobao simulator provides a "live" environment, where the agent can be tested with the virtual customers and recommendation system. Based on the query of the customer, the system returns a good list, and the virtual customer will decide whether to click the items in the list. The virtual customer will change interest after each interaction to simulate the real world situation.
% Each virtual customer has 3-dimensional dynamic attributes and 11 static attributes, such as age and gender, which is encoded into 88 binary dimensions. The attributes of each item are encoded into a 27-dimensional space.

\vspace{1mm}\noindent\textbf{Baselines.}
Several state-of-the-art counterfactual data-augmented sequential recommendation methods are not open source and are not deployed in a comparable environment with ours. Therefore, we deploy the proposed Counterfactual Synthesis Policy into various popular Reinforcement Learning algorithms and compare them with original algorithms to evaluate the efficiency of the proposed method. Specifically, the Counterfactual Synthesis Policy is compared with the following baselines:

\begin{itemize}
    \item \textbf{Deep Deterministic Policy Gradient (DDPG)}~\cite{lillicrap2015continuous}: A model-free and off-policy algorithm for environments with continuous action spaces.
    \item \textbf{Soft Actor Critic (SAC)}~\cite{haarnoja2018soft}:  An off-policy maximum entropy Deep Reinforcement Learning algorithm with a stochastic policy.
    % \item \textbf{Truncated Quantile Critics (TQC)}~\cite{kuznetsov2020controlling}: An algorithm alleviates the overestimation bias by making use of quantile regression to predict a distribution for the value function.
    \item \textbf{Twin Delayed DDPG (TD3)}~\cite{fujimoto2018addressing}: An algorithm that improved performance over baseline DDPG by introducing three critical tricks, including learning two Q-functions instead of one, updating the policy less frequently and adding noise to the target action.
\end{itemize}

% \vspace{1mm}\noindent\textbf{Evaluation Metric.}
% We employ click-through-rate as the numerical criteria for evaluation. Since each episode may end up with different number of steps, we computed the click-through-rate as:
% \begin{equation}
% \label{eqt-l-inf}
% \begin{aligned}
% CTR = \frac{r_{episode}}{10*N}
% \end{aligned}
% \end{equation}
% where $r_{episode}$ is the reward for a single episode and $N$ is the number of steps in the episode. 10 indicates that the customer has clicked all top 10 items with the highest values recommended by the system. Higher values of CTR indicate better performance.

\vspace{1mm}\noindent\textbf{Implementation Details}.
We use DDPG to train our counterfactual synthesis policy with the parameters: $\gamma = 0.95, \tau = 0.001$, learning rate of 0.003 and the hidden size of 128. We train the policy for 1,000,000 episodes with mini-batch size 5. 

After training the counterfactual synthesis policy, we deploy our policy into reinforcement learning algorithms. The brief illustration is depicted in the right part of Figure 2. 
To be specific, when the training agent interacts with the environment in the actor network and stores the transition $(s_t, a_t, s_{t+1}, r_t)$ into the replay memory buffer, the agent with our trained policy will also take action and change the environment from $s_t$ to $s_{c, t+1}$. The obtained next state $s_{c, t+1}$ is a counterfactual state for the original state $s_t$, of which the trivial components has been replaced. To expanding the replay memory buffer, let the training agent interact with the environment under counterfactual state $s_{c, t+1}$ and store the transition $(s_{c, t+1}, a'_t, s_{t+1}, r'_{t+1})$ into replay memory buffer.
The model is carried out using PyTorch, and the experiments are conducted on a server that consists of two Intel (R) Xeon (R) CPU E5-2697 v2 CPUs, 6 NVIDIA TITAN X Pascal GPUs and 2 NVIDIA TITAN RTX.

\subsection{Overall Comparison}
The overall comparison result is reported in Figure 2(a). In a nutshell, we observe that our approach can achieve a considerable improvement over the selected baselines.
Among the baselines, DDPG achieves the best performance while taking longer than others.
SAC suffers significant variance, and the possible reason would be that SAC utilizes a stochastic policy that introduces extra noise to the agent. TD3 does not perform well as expected but has a lower variance than DDPG and SAC. The bad performance may be caused by the delayed update parameter mechanism.
For DDPG, deploying the counterfactual synthesis policy shows a clear improvement after 45000 episodes and enhances DDPG by around 43.86\% in CTR. As can be seen from the figure, the performance of DDPG with our approach is more stable late during the evaluation.
For SAC, our policy enhances its performance after 25000 episodes by up to 56.7\% but does not alleviate the significant variance problem. With regard to TD3, our policy also shows an improvement of around 48.9\% but is not very stable in the early stage of the evaluation. During the later stage, the performance of TD3 employing our policy is more stable than itself.
We also performed a comparison of the average reward per episode for the three baselines with or without our policy. The results in Table 1 show that our policy improves selected baselines considerably. However, the remarkable thing is that the significant improvement over TD3 is the result of its lower base. And the possible reason for the pool reward of TD3 is that it would probably explore enough to find helpful learning signals.
% \begin{figure}[h]
%   \centering
%   \includegraphics[width=\linewidth]{fig2.png}
%   \caption{Overall comparison results}
%   \Description{.}
% \end{figure}

\begin{figure}
     \centering
     \begin{subfigure}[b]{0.48\linewidth}
         \centering
         \includegraphics[width=\linewidth]{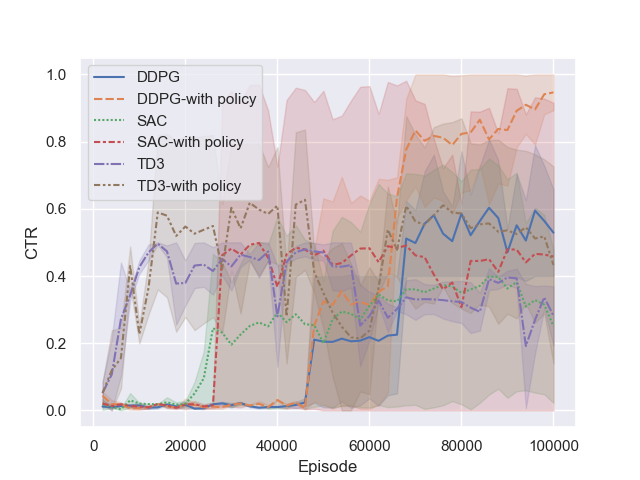}
         \caption{}
     \end{subfigure}
     \begin{subfigure}[b]{0.48\linewidth}
         \centering
         \includegraphics[width=\linewidth]{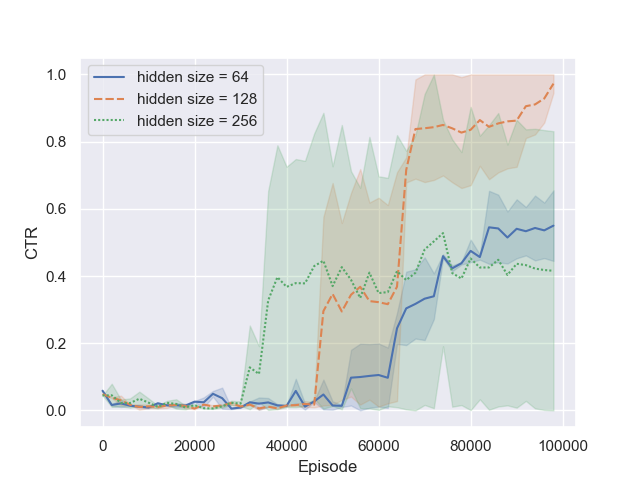}
         \caption{}
     \end{subfigure}
        \caption{(a): Overall comparison result between the baselines and baselines employed our policy. (b): Performance comparison on the hidden size of 64, 128 and 256.}
\end{figure}

\begin{table}
  \caption{Average reward per episode}
  \label{tab:freq}
  \begin{tabular}{ccc|c}
    \toprule
    Baseline &Original&With our policy&Improvement\\
    \midrule
    DDPG & 17.273681 & 28.398414 & 64.40\%\\
    SAC & 18.372824 & 20.292473 & 10.45\%\\
    TD3 & 6.027942 & 24.435 & 305.36\%\\
  \bottomrule
\end{tabular}
\end{table}

\subsection{Hyper-parameter Study}
% \begin{figure}
%     \centering
%     \includegraphics[width=\linewidth]{fig3.png}
%     \caption{Performance comparison on the hidden size of 64, 128 and 256.}
%     \label{fig:my_label}
% \end{figure}

We analyze the number of hidden sizes to investigate their influence on the learning of counterfactual synthesis policy. We set the hidden size number in the range $[64, 128, 256]$, and present the comparison result in Figure 2(b). The figure shows that the best performance is achieved when the hidden size is the moderate number 128. Either too small or too large number would drop the performance of our counterfactual synthesis policy. Although the standard deviation of the result with 64 hidden sizes is small, it has a slower convergence speed and worse performance, which may attribute to too few neurons in the hidden layers to collect information adequately. A too large number of hidden sizes results in a large standard deviation. One possible reason is that using too many neurons in the hidden layers result in an overfitting problem making it hard to converge.

\section{Related Works}
\vspace{1mm}\noindent\textbf{RL-based interactive recommendation.}
Since RL trains an agent that is able to learn from interaction trajectories, it has shown great effectiveness in modelling interactive recommendation processes.
~\citet{mahmood2007learning} modify the Markov Decision Process (MDP) to achieve an adaptive interaction recommender system using Reinforcement Learning (RL) techniques.
~\citet{zheng2018drn} apply a DQN structure and Dueling Bandit Gradient Descent method in reinforcement learning framework to address the dynamic nature of news recommendation.
To handle the drawback of DQN in large and dynamic action space scenarios, ~\citet{zhao2017deep} build the list-wide recommendation model with the capability of dealing with large and dynamic item space upon the Actor-Critic framework.
~\citet{chen2020knowledge} utilize the knowledge graphs on the actor-critic network to improve the stability and quality of the critic network for interactive recommendation.

\vspace{1mm}\noindent\textbf{Counterfactual for Recommendation.}
Causality and counterfactual reasoning attracted increasing attention in many domains. 
Many researchers leverage counterfactual reasoning in recommendation system to release bias issues.~\citet{schnabel2016recommendations} propose a framework to handle selection bias in the learning of recommendation systems by adapting techniques from causal inference. 
~\citet{liu2020general} solving the general bias problems in a recommendation system via uniform data. 
Some works focus on making a connection between classical recommendation models and causality to correct the exposure bias~\cite{bonner2018causal, wang2020causal}. 
A growing number of works take the insights of counterfactual reasoning to provide explanations, increase the interpretability of the model and learn a robust representation~\cite{tan2021counterfactual, madumal2020explainable, zhang2021causerec}. ~\citet{zhang2021causerec} use similarity function to identify the subsets in the behaviour sequence that can be replaced to generate counterfactual sequences, which help to learn a robust user representation for sequential recommendation.
Another line of work focuses on data-augmenting of the training samples for data-scarce tasks.
~\citet{wang2021counterfactual} propose a counterfactual generation framework based on the embedding space for sequential recommendation models. 
~\citet{lu2020sample} focus on the learning of a causal mechanism in the counterfactual generation process for data augmentation.

\section{Conclusion}
In this paper, a novel counterfactual synthesis policy has been proposed to generate counterfactual interactive data automatically.
We employ the DDPG structure to train our policy to handle the data sparsity problem in RL-based interactive recommendation systems. 
% Our proposed policy has the advantage of taking the dynamic preferences of users into account. 
Furthermore, the counterfactual synthesis policy can easily be deployed in different RL frameworks to generate counterfactual data in the training process.
Experiment results show that counterfactual synthesis policy performs well on different RL frameworks and achieve a considerable improvement for all these baselines.
As for future works, we are planning to make more exploration to merge the training process of the target agent and counterfactual synthesis policy. 

% In addition, there exists the situation that essence components have been replaced by counterfactual synthesis policy. Therefore, optimizing the identification of essence and trivial components is also a direction of our future work.

\bibliographystyle{ACM-Reference-Format}
\bibliography{sample-base}
\end{document}